\documentclass[sigconf]{acmart}
\usepackage[listings, many]{tcolorbox}

\AtBeginDocument{%
  }

\setcopyright{none}
\usepackage{booktabs}
\usepackage{array}
\usepackage{booktabs}
\usepackage{array}
\usepackage{tabularx}
\copyrightyear{2026}
\acmYear{2026}
\setcopyright{cc}
\setcctype{by}
\acmConference[ASSETS '26]{The 28th International ACM SIGACCESS Conference on Computers and Accessibility}{October 25--28, 2026}{Vila Nova de Gaia, Portugal}
\acmBooktitle{The 28th International ACM SIGACCESS Conference on Computers and Accessibility (ASSETS '26), October 25--28, 2026, Vila Nova de Gaia, Portugal}
\acmDOI{10.1145/3797867.3832767}
\acmISBN{979-8-4007-2521-0/2026/10}
\usepackage{ifthen}
\usepackage{xspace}
\usepackage{etoolbox}
\usepackage{cleveref}
\usepackage{paralist}
\usepackage{enumitem}
\usepackage{xcolor}
\usepackage{soul}
\newtoggle{comments} 
\toggletrue{comments}
\newcommand{\etal}{%
   \emph{et al.}
}
\begin{document}
\raggedbottom

\title{Reflections and Recommendations on AI Adoption Practice from a Mixed-Ability Research Group}
\author{Shalini Madan}
\email{shalinii@umich.edu}
\affiliation{%
  \institution{University of Michigan}
  \city{Ann Arbor}
  \state{Michigan}
  \country{USA}
}

\author{Sreelakshmi Surabiyil Bindu}
\email{sreelaks@umich.edu}
\affiliation{%
   \institution{University of Michigan}
  \city{Ann Arbor}
  \state{Michigan}
  \country{USA}
}
\authornotemark[1]
\author{Veronica Pimenova}
\email{pimenova@umich.edu}
\affiliation{%
   \institution{University of Michigan}
  \city{Ann Arbor}
  \state{Michigan}
  \country{USA}
}
\authornotemark[1]
\author{Ellie Seehorn}
\email{seehorn@umich.edu}
\affiliation{%
  \institution{University of Michigan}        
  \city{Ann Arbor}
  \state{Michigan}
  \country{USA}
  }
\authornote{The second, third, and fourth authors contributed equally to this work.}

\author{Venkatesh Potluri}
\email{potluriv@umich.edu}
\affiliation{%
  \institution{University of Michigan}        
  \city{Ann Arbor}
  \state{Michigan}
  \country{USA}
}

\begin{abstract}
Generative AI tools have recently been rapidly adopted by academics in mixed-ability research teams for both personal and professional tasks. While previous work on adoption of AI-based workflows has focused on collaboration and productivity, the perceptions of AI use within research teams remains divided. Through qualitative analysis of interviews of the five members of our mixed-ability research team, we discuss the motivations, challenges, and practices surrounding the use of generative AI in our lab. We reflect on experiences that shaped recommendations for balanced AI use that enable mixed-ability team workflows: (1) managing disability tax \& crip time, (2) homogenizing identity, (3) risk disclosure of private information, (4) self-experimentation and miscellaneous tasks, and (5) information seeking. We build upon these themes to present AI practice recommendations we established for our lab to promote AI workflow adoption while preserving agency and disability identity.
\end{abstract}



\maketitle
\section{Introduction \& Related Work}
\label{ssec:Intro}
The proliferation of generative AI through dedicated apps and features built into consumer-facing tools is causing a significant spike in the adoption of artificial intelligence (AI) across a range of tasks~\cite{wang2024understandinguserexperiencelarge,huang2026want,costagomes2025temporal}. An estimated 43 percent of U.S based knowledge workers use generative AI multiple times a week, with ChatGPT, a popular AI chatbot having 800 million weekly active users~\cite{Chatterji2025HowPeople}. Generative AI is positioned as a tool that significantly changes how we perform professional tasks such as coding, research, and writing, and personal tasks such as cooking, searching for information, and content creation~\cite{zheng2024disciplink,ning2025aroma, lyu2024preliminary,Adnin:2024:Kingofknowledge}.
Studies discuss the role of AI agents in workplace productivity, collaborative settings~\cite{johnson:2025:aicollab, lee_genAI_critical_thinking, zhang_ui_prototype}, and in enabling mixed-ability team collaboration~\cite{xiao2026:mixedabilitycollab}. Findings in Weiz~\etal~\cite{weisz:2025:examining} note that AI agents in enterprise settings could serve as junior developers, producing high-quality code with faster turnaround times. Jang~\etal~\cite{Jang:2024:LLMs} discuss the benefits of LLMs for autistic employee communication in workplace settings due to AI's ability to organize information and respond sooner to queries. 

Added to this, findings from many studies discuss the potential of Generative AI to bridge and assist in managing various access needs~\cite{glazko:2025:generative,Mullen:2024:AI}. Glazko~\etal~\cite{Glazko:2023:GAIAutoEthno} provide an earliest account of the value of generative AI to increase accessibility. They note AI's capability to assist with visual imagery and making content accessible. For blind or low-vision (BLV) programmers, these generative AI tools improved accessibility by reducing barriers in programming that relied on visual ability~\cite{cha2025game,Adnin:2024:Kingofknowledge} and made hands-free visual assistance possible~\cite{huh:2025:uist}. These tools offered new ways of self-expression for people with disabilities that impact communication~\cite{Mo:PhD:2026}. Increased AI capabilities have also enabled new, commercially available technologies to access images~\cite{bemyai,AIRA}, use inaccessible interfaces, and work~\cite{BeMyAIwork}.

While promising, challenges such as hallucinations, misinformation, ableism, bias, and privacy concerns in Human-AI interaction persist~\cite{tang2025everyday,phutane:2025:ableism, Kameswaran:2026:aihiring,alharbi:2024:misfitting}. For example, Weisz~\etal~\cite{weisz:2025:examining} discuss how employees in enterprises note that AI agents still require human supervision and expertise to review AI-generated content. In addition, AI adoption in workplace settings poses challenges such as employee burnout, cognitive decline, and loss of creative thinking, calling for organizations to outline rules for AI practice~\cite{ranganathan2026ai,weisz:2025:examining,li:2024:user}. Li~\etal~\cite{li:2024:user} discuss concerns about AI accuracy and its implications for User Experience (UX) designers' reasoning, which ultimately pose bigger challenges in UX team setups, such as creativity exhaustion from keeping up with the pace of GenAI outputs. Similarly, Johnson~\etal~\cite{johnson:2025:aicollab} highlight the need for responsible AI practices across organizations to prioritize psychological safety in AI-enabled workplaces in order to protect employee privacy and well-being.

The variability in benefits, risks, access, and use of AI in organizations raises a common concern in determining both the context and the extent of reliance on AI integration into daily tasks. In response, this experience report discusses the various motivations for using or rejecting AI in mixed-ability teams in academia and examines AI's role in a world built for humans, yet failing to address the needs of all humans equally—the very reason disability is disabling. We contextualize the use of AI for both personal and research purposes through qualitative findings from five semi-structured interviews with team members, followed by thematic analysis~\cite{guest2011applied}. Through this synthesis, \textbf{we contribute AI Practice guidelines for our research lab}, in anticipation that this contribution could guide other academic research labs with mixed-ability teams attempting to balance the advantages and risks of AI adoption.
\section{Team and Methodology}
We introduce our team through a positionality statement, three user personas, and the data collection process we used during the writing and development of this experience report. 

\subsection{Positionality}
We are a research team of incoming and early-stage Ph.D. Students, accessibility support staff, and a tenure-track Assistant Professor in an R1 Research University in the United States. Our mixed-ability group has contributed to programming environments that increase accessibility, auto-ethnographic accounts of technology and AI, and guidelines to inform the design of generative AI experiences. Team members have 1 -- 8 years of research experience, with most members having less than 5 years. With programming being a popular domain impacted by AI (e.g. "vibe coding" ~\cite{sarkar2025vibe, qual_co_creation_vibe_coding, vibe_code_practices_challenges}), and the potential for AI to address our accessibility needs, our group often engages in discussions that encourage critical and pragmatic perspectives about AI adoption and use. Outside of research contexts, the team would often engage in informal conversations about AI, building accessible versions of inaccessible tools for the lab with AI, and testing and iterating on them over time.

\subsection{Personas}
To respect the disability disclosure preferences of the authors, we surface AI use experiences and quotes through aggregate personas representing the diverse abilities and experiences of each author.

\paragraph{Persona 1:} Taylor is a graduate student with two years of domain knowledge in their field of research and identifies as neurodivergent. Their accessibility needs include extended deadlines and requests for flexible meeting modes (virtual/in-person). Taylor uses AI for tasks they already know how to do, such as vibe coding using tools such as GitHub Copilot and Replit, and for personal tasks such as communication and time management. However, Taylor refuses to use AI for any work that involves creating new knowledge (e.g., writing), as they believe that using AI for such tasks will affect their ability to learn how to disseminate information due to its monotonous tone and predictable writing style. 

\paragraph{Persona 2:} Evan is a research professional with adequate domain knowledge who does not identify as having accessibility needs. Evan uses AI for tasks such as programming and other domain-relevant sub-tasks. Evan always checks AI-generated output and has thorough knowledge of any topic for which they use AI. Evan's biggest concern is privacy and disclosure of information to AI agents. Evan believes there is a dire need for AI users to be able to opt in or out of where their data is used, something that not all platforms make clear.

\paragraph{Persona 3:} Jay is a seasoned researcher with over eight years of domain knowledge and a tenure-track professor at an R1 university who identifies as a blind person. Jay relies on screen readers and other assistive technologies and uses AI tools such as Seeing AI, Be My AI, and Claude to support accessibility, administrative work, calendaring, and information management in their daily life. Jay believes that over-reliance on AI is not beneficial as it may affect the ability to maintain critical skills and preserve unique perspectives for task completion.

\subsection{Data Collection}
The informal ideation of this project occurred during a discussion and brainstorming session in early March about the over-reliance on AI~\cite{ranganathan2026ai} and the potential for these tools to cause burnout~\cite{bedard2026using}, when Taylor commented that ``\emph{I have a college degree, so I should be able to write an email}''. Evan expressed similar sentiments, and these discussions revealed the nuanced considerations the team members were making in their AI adoption. Notably, all team members paused and reflected on AI use for some tasks, suggesting that friction is an adopted practice. Participation in this experience report was optional, and all steps of self-reflection were emphasized to meet the disclosure preferences of the team members.

Our data collection was both prospective and retrospective. We use a semi-structured interview protocol, memos, and thematic analysis as tools to guide our process, and not as rigorous study instruments. To facilitate the prospective portion of our data collection, the team members discussed the goal of the project, and one author designed an initial semi-structured interview protocol. The group collectively reviewed the protocol, modified questions, and added new questions. This protocol captured reflections on AI use cases, including accessibility, motivations for adding friction, experiences where AI succeeded and failed, and external factors. 

Each team member interviewed one other member, and Jay did not act as an interviewer to avoid the potential for power dynamics influencing the data. Interviews were audio- and video-recorded, and each interviewer memoed~\cite{birks2008memoing} the interview afterward. The retrospective portion of our data collection involved the capture of AI use with guiding questions asking about the tasks that AI was used for, motivations, and to identify if there were other tasks that AI could support in the future. Team members manually reviewed chat histories, used prompts asking AI models to show``\emph{highlights from our recent conversations}'' and responded to these prompts in a shared Google Doc. To identify recurring themes, the group held two meetings and reflected between discussions. Subsequently, all authors refined these themes and developed a set of AI practice recommendations through multiple rounds of discussion. 

We recognize that the guidelines we propose are specific to our research lab and may not be generalizable to all research groups. However, it is worth noting that the nature of accessibility barriers, expectations, and circumstances that informed these guidelines are broadly applicable to academia ~\cite{jain:2020:navigating,hammer:2020:lab,Ezeamii:2025:stem,shinohara:2020:Inequitable}. Our human-centered process to establish our lab guidelines holds a broader potential for other mixed-ability collaboration groups.

\section{AI Practice Recommendations}
\label{ssec: results}
Our findings reveal five core themes reflecting our mixed-ability team’s views on AI. We identify its potential to manage disability tax, surface implications to privacy and disability disclosure, refuse AI use to preserve novelty, and use it for experimentation and information seeking. We present recommendations for AI practice developed from this synthesis.   
\subsection{AI to Manage  Disability Tax \& Crip Time}
In an academic environment, multiple tasks such as conference submissions, grant deadlines, reviewing, and mentoring run on tight timelines. Additionally, administrative and paperwork for various research-related activities must be completed within a short timeframe, which adds to the workload. We reflect that these tasks and related expectations are often grounded in ableist norms and rely on inaccessible tools, offering little to no flexibility for disability-specific flare-ups or for additional time one may need when using accessibility workarounds to complete them. Our findings indicate that AI may help with planning, scheduling, and completing high-effort, low-priority ``busy work'' tasks. For example, Jay notes that using AI for proxy communication (drafting emails) could help reduce the time spent coordinating with human support, where an accessible interface should have been the norm:

\begin{tcolorbox}[breakable,colback=blue!5!white,colframe=blue!50!black, title=Jay on AI for Communication]
\textit{``I feel really bad for considering the option to use a bot to talk to a human. But part of my motivation comes from anger, quite frankly; if my important time can be spent navigating accessibility barriers that shouldn’t exist, I also wouldn’t be hesitant to proxy-communicate in some of these cases.''}
\end{tcolorbox}

Similarly, Taylor notes that AI has been helpful while managing scheduling and emailing tasks during disability flare-ups:

\begin{tcolorbox}[breakable,colback=pink!10!white,colframe=red!50!black, title=Taylor on AI for Schedule Management]
\textit{``\ldots If I do have an episode of chronic pain, and I can't come to an event, then it's easier for me to have a calendar to see what I'm supposed to be at, and then I can click the delete the event, or reschedule it for a later time\ldots I do use the AI features within Google Calendar that already exist, for naming an event, or rescheduling an event, things like that\ldots And then also for sending myself email reminders, I sometimes use  the AI feature where it will complete the emails where I'll start writing it and then it'll finish the text for me.''}
\end{tcolorbox}

AI surfaced as a means to manage time for team members outside of accessibility contexts. Evan notes that AI has helped in situations when they were pressed for time:

\begin{tcolorbox}[breakable,colback=green!4!white,colframe=green!50!black, title=Evan on AI for Time Management]
\textit{``\ldots When I'm short on time and I really need an expert\ldots I think under situations where I really need a quick answer, that's when I would turn to AI.''}
\end{tcolorbox}

In our experience, using AI for high-effort, low-impact tasks helps reduce disability tax and crip time~\cite{jamshed:2025:ai}, making time work for us and reducing instances of working with timelines that do not account for the uncertainties that arise with accessibility needs.
\paragraph{\textbf{AI Practice Recommendations:}}
\begin{itemize}
    \item \textbf{R1}: Consider using AI to plan organizational tasks or busy work by identifying opportunities where accessibility issues in the environment or disability-related flare-ups can shift how time works, as AI can remove the extra time a person would otherwise need to complete a task. We argue that addressing systemic barriers to accessibility should continue to be the best practice, and we do not offer this recommendation as a replacement. Expectation to identify these opportunities can place the burden of access on people with disabilities. Instead, we offer this as a strategy in cases where infrastructures and ableist practices fail people with disabilities.
\end{itemize}

\subsection{AI Homogenizes Identity}
Our findings indicate that all team members collectively refuse to use AI for research, creative tasks (e.g., design), or other professional or personal tasks, which may undermine the novelty or uniqueness of one's writing or creative style, as AI would introduce monotony and potentially take away the joy of creating artifacts. Additionally, team members who identify as researchers with disabilities refused the use of AI for tasks that remove their unique tone or perspective, a core component of their identity. Taylor notes: 

\begin{tcolorbox}
[breakable,colback=pink!10!white,colframe=red!50!black, title=Taylor on AI and Novelty]
\textit{``I'm in academia, right? I spent a very long time learning how to think, and it seems like that's a pretty important skill to just let go by the wayside. I don't actually use it (AI) very much for writing\ldots it's not gonna degrade my ability to do anything the first time I try it. But, at some point, if you develop a habit of using AI, where are my contributions in this? And that's something that I still want to uphold.''}
\end{tcolorbox}

Evan noted that AI is not quite ready to replace a researcher's core competency yet, where they describe how their own experience with AI tools for qualitative analysis could not replace a researcher's own analytical depth:

\begin{tcolorbox}[breakable,colback=green!4!white,colframe=green!50!black, title=Evan on AI and Ability]
\textit{``A lot of people talk about how AI is going to replace all of us and it's going to do a lot of things\ldots I don't think that's true because of my own experiences. We were introduced to a tool that gave summaries of the qualitative data. The summaries and the inferences that AI made were nowhere close to what I would have made as a qualitative researcher\ldots I don't think [AI] has the ability to find insights like qualitative feedback}\ldots One skill that I wouldn't see AI do for me is the ability to identify minor details in qualitative research that can infer a lot of things\ldots research synthesis, I think is one skill of mine that would probably not be affected by AI.''
\end{tcolorbox}

We find the use of AI for time management, scheduling, and tool-building beneficial for mixed-ability teams such as ours. However, we note the need to define a baseline to balance cognitive capability and AI assistance for task completion. Jay discusses trade-offs and their thoughts when making a decision to use AI for time-sensitive decision-making that is impacted by accessibility barriers: 

\begin{tcolorbox}[breakable,colback=blue!5!white,colframe=blue!50!black, title=Jay on AI and Barriers to Access]
\textit{``\ldots Carefully thinking about accessibility and time, especially if we had a mixed ability team, right? Time is a concept that I kind of think about in the context of AI\ldots? It has a lot of potential to reduce the disability tax that we have to pay in some cases. But I think framing that really, really carefully and thinking about\ldots whether it is actually reducing my accessibility barriers or is it actually reducing my cognitive strength as a researcher\ldots''}
\end{tcolorbox}

Jay notes that a mixed-ability team setting can sometimes reduce accessibility barriers. However, Jay still calls for a clear baseline process to guide AI use and avoid over-reliance:

\begin{tcolorbox}[breakable,colback=blue!5!white,colframe=blue!50!black, title=Jay on AI and Mixed-ability Teams]
\textit{``Yes, I'm balancing other things, but in some ways, [the team] amplifies my productivity in some ways or the other. We are all working on projects together, so I think I'm very aware of that\ldots It does come with a little bit of a privilege. However, I still think that these baseline recommendations will still hold, especially with accessibility, because there can be a time crunch, and AI can look like a very tempting solution\ldots''}
\end{tcolorbox}
 
Across findings, we note that team members also agreed to use AI only to complete a specific set of tasks (e.g., making website updates, vibe-coding) rather than to build skills relevant to their core competencies. 
\paragraph{\textbf{AI Practice Recommendations:}} 
\begin{itemize}
\item \textbf{R2}: Avoid using AI for tasks that risk the expression of your unique identity, creativity, and authenticity. In the context of academia, research communication, or writing, disability identity is valuable and offers a new perspective that cannot be replicated by AI-generated monotone writing~\cite{Agnew:2024:IllusionArtificialInclusion,kosmyna2025brainchatgptaccumulationcognitive, storey2026technical}. 
\item \textbf{R3}: Identify a set of core competencies for yourself and with your mentors. Consider avoiding AI use for acquiring, retaining, and demonstrating core competencies. Organizations can differentiate between skill-building and task completion, even though these are not mutually exclusive, by setting expectations for specific tasks. AI could be potentially used to complete a task, and may be avoided in cases where the core goal is to develop a skill by completing the task. For example, AI coding tools could be used to update a lab website. However, the use of AI may not be appropriate to develop code for data analysis, especially if the core goal for a researcher is to acquire the skill, as over-reliance on AI could result in over-delegation and inability to identify errors~\cite{obrien2026surveygenerativeaiadoption}.
\end{itemize}
\subsection{AI Risks Disclosure of Private Information} 
We observe that all team members were skeptical about how AI protects against the disclosure of disability or other sensitive identifying information they may have previously shared. In larger lab or organizational settings, unclear norms around disability disclosure could risk bias and create an environment that is not inclusive. All team members collectively reflected on the need for well-defined guidelines to protect identity and disability information when interacting with AI. Jay notes that:

\begin{tcolorbox}[breakable,colback=blue!5!white,colframe=blue!50!black, title=Jay on disclosure to AI]

\textit{``I think for AI, there are a few considerations I make\ldots Am I giving up the privacy of my own self beyond what I'm comfortable with and maybe compromising the privacy of others, that is one case\ldots even with Be My Eyes, I don't give it photos of some people who I know are very privacy minded In my life. \ldots even for things like Claude and so on, I am careful. I'm very tempted to connect Gemini, for example, to Google Workspace, because I think I'll find a lot of value as somebody who has to sift through a lot of information. It will add value. However, the way Google has its control is all or nothing, right? So I'm like, I don't want to give it access to my email where I have student data, or I'll have study data in my Google Drive\ldots There are very few things that I want to give access to, and I don't have that granular control. So yeah, privacy is one aspect\ldots''}
\end{tcolorbox}

Our discussions also surfaced another tension: the possibility of disclosing a teammate's disability to make content accessible, or in a potential use case where AI reads unpublished drafts of auto-ethnographic accounts or assists with formatting reference letters that may contain sensitive information. 
\paragraph{\textbf{AI Practice Recommendations:}}
\begin{itemize}
    \item \textbf{R4}: Mixed ability teams must introduce communication norms on de-anonymization of any team member's disability information while communicating with humans and AI.  
    \item \textbf{R5}: Organizations should communicate data privacy policies. When AI tools are made available at an organizational level, clear guidelines should be communicated for allowed and prohibited use cases. These could be developed in consultation with experts with disabilities, as needs vary~\cite{binns2021could}.
    \item \textbf{R6}: Check in with team members about their disclosure preferences before using team/shared AI accounts to promote individual privacy/autonomy and prevent disability bias. 
\end{itemize}
\subsection{AI for Self Experimentation and Miscellaneous Tasks}
We observed that all team members express curiosity to explore AI capabilities across personal interests, such as cooking and astrology, as well as skill-based experiments, such as tool building with AI. Findings show that personal use by some team members may have improved their understanding of AI and its applicability to professional use cases. For example, Evan noted that they use AI for creating recipes, exploring planets, and astrology. Taylor noted the use of AI for website development and design, and, informed by his personal experiments, Jay led one of the lab meetings where the team vibe-coded an accessible "when2meet" tool, and, informed by his personal experiments, also designed their own qualitative analysis tool, Qua11y. While we discuss the potential risks of over-reliance on AI, the team collectively agrees that experimenting with AI could often benefit their professional development and promote knowledge/skill transfer among the team. To promote skill-building through experimentation, Jay introduced fixed annual funding for each lab member with the purpose of exploring and experimenting with AI tools, including use cases such as vibe coding.
\paragraph{\textbf{AI Practice Recommendations}}
\begin{itemize}
 \item \textbf{R7}: Make space for self-experimentation, even in personal use or non-professional use. Create spaces for sharing without judgment. Consider allocating 10\% of your time to areas where AI could be helpful, and organize a quarterly skills-sharing meeting to discuss your experience with AI use. 
 \item\textbf{R8}: In academia, lab leaders could explore a fixed no-approval budget for exploratory AI tool usage. 
\end{itemize}
\subsection{AI for Information Seeking}
We previously noted that the team collectively refused to use AI for skill-building or for tasks requiring the application of their core competencies. However, in contexts where the team already has sufficient domain knowledge to verify information generated by AI, we observe that the team collectively agreed to use AI for low-risk tasks, such as information seeking or rapid prototyping/programming. For example, Evan highlights the requirement to have the skills to verify AI-generated information: 

\begin{tcolorbox}[breakable,colback=green!4!white,colframe=green!50!black, title=Evan on Skills for Verification of AI Output]

\textit{``You don't have to assume that it (AI) is going to become perfect\ldots I feel like skills will still matter, because, like I said, it won't be 100\% perfect, so\ldots you still need to know how to deal with the stuff that it doesn't get right, or you need that knowledge to actually know whether what it's saying is correct or not. Skills will still be important in everything.''}
\end{tcolorbox}
Similarly, Taylor notes that they use AI only when they can verify the code it generates while building tools for low-stakes tasks. Jay, who uses AI to build tools, reflected on a past experience in which they spent several hours iterating with AI, ultimately leading to burnout from excessive AI use. Jay also notes that while the process of building with AI can be exciting, if there are no defined boundaries on when to stop using AI or to stop iterating, the risks of AI burnout and loss of time can be high, ultimately limiting one's productivity. Teammates introduced friction by asking more questions about a task to gain additional context or by intentionally using AI tools that do not integrate with workflows in which certain tasks are performed.
\paragraph{\textbf{AI Practice Recommendations}}
\begin{itemize}
    \item\textbf{R9:} Always verify information generated by AI. One way to do so is by introducing friction in tasks by designing workflows that support asking questions, or by using tools that do not tightly integrate into the workflows for certain tasks that necessitate friction.
    \item\textbf{R10:} Recent work notes that AI could intensify workload and productivity~\cite{bedard2026using}. To avoid burnout, time-block AI use for specific tasks to limit cognitive effort.
\end{itemize}

\section{Discussion}
\label{ssec:Result}
 We observe a divide in AI use or refusal in academia, related to recent work (e.g. ~\cite{nelson2025academic, pinzolits2024ai, beshr2024ai}). For example, Mann~\etal ~\cite{mann2026aifutureacademicpeer} discuss the skepticism about adopting AI due to several risks, such as hallucinations, data leaks, and threats to critical reasoning. However, they argue that defining clear guidelines to protect privacy, novelty, and autonomy could resolve the challenges of AI use. Further, AI is also gaining the perception of being effective for tasks like information gathering within academic research~\cite{chubb2022speeding}. Contrastingly, we observe a stronger inclination of AI use in industry organizations,  encouraging employees to adopt AI in their everyday workflows~\cite{weiss2025nvidia, marr2025microsoft}. In some cases, organizations are aligning incentives with higher token usage and generated lines of code without the necessary systems in place to differentiate experimentation from work, leading to newer kinds of technical and cognitive overload that some are calling the~\emph{AI brain fry}~\cite{storey2026technical,zhang2025technicaldebtaiassisteddevelopment,bedard2026using}.

Across these varied contexts, however, the adoption of generative AI tools raises tensions that require balancing perceived productivity with employee well-being and human judgment ~\cite{pm_genAI, AI_devs_daily}. Developing AI practices is one way of promoting practical and meaningful adoption of AI while addressing these concerns~\cite{ranganathan2026ai}. Our recommendations could support a more sustainable adoption of AI tools while making space for self-experimentation. Further, many conversations guiding AI adoption have not given significant consideration to accessibility-specific use cases. Accessibility studies in workplace contexts focus on impacts assuming adoption~\cite{Adnin:2024:Kingofknowledge,chen2025screenreaderprogrammersvibe,alshaigy:2024:forgotten}. Our work offers a complementary perspective of recommendations that are developed through the adoption of human-centered research methods, accessibility needs, and self-experimentation. In doing so, we add to the body of knowledge of AI policies that reflect the values, preferences, and personal experiences of mixed-ability teams-- an identified need evident from our own experience, and can serve as an example for meaningful adoption across similar mixed-ability labs in academia. Further, our guidelines show potential towards organizational-level adoption, given that relevant team-specific contextualized modifications are made to our recommendations. We hope that this experience report inspires further nuanced conversations on AI use and leads to the development of human-centric, accessibility-first AI practice recommendations for the community at large.
\bibliographystyle{ACM-Reference-Format}
\bibliography{references}
\end{document}